\documentclass[sigconf]{acmart}
\usepackage{multirow}
\usepackage{balance}
\usepackage{caption}
\usepackage{placeins}
\usepackage{subcaption}
\usepackage{amsfonts}       
\usepackage{nicefrac}       
\usepackage{microtype}      

\AtBeginDocument{%
  \providecommand\BibTeX{{%
    \normalfont B\kern-0.5em{\scshape i\kern-0.25em b}\kern-0.8em\TeX}}}

\newif\ifNoComments
\NoCommentstrue
\newif\ifProduction
\Productiontrue

\newcommand{\ttt}{\texttt}

\newcommand{\rqone}{\textbf{RQ1}}
\newcommand{\rqtwo}{\textbf{RQ2}}
\newcommand{\rqthree}{\textbf{RQ3}}
\newcommand{\rqfour}{\textbf{RQ4}}
\newcommand{\rqfive}{\textbf{RQ5}}
\newcommand{\ststr}{\phantom{$^*$}}

\usepackage{color}

\ifProduction
\newcommand{\anonref}[1]{\cite{#1}}
\newcommand{\flexneuart}{FlexNeuART \cite{FlexNeuART}}
\NoCommentstrue
\else
\newcommand{\anonref}[1]{[\textbf{anonymous}]}
\newcommand{\flexneuart}{an \textbf{anonymous} retrieval toolkit}
\fi

\ifNoComments
\newcommand{\mynote}[1]{}

\newcommand{\pb}[1]{}
\newcommand{\lb}[1]{}
\newcommand{\ym}[1]{}
\else
\newcommand{\mynote}[1]{\textbf{\color{red}\small #1}}

\newcommand{\pb}[1]{\textcolor{green}{{\textbf{PB:} #1}}}
\newcommand{\lb}[1]{\textcolor{orange}{{\textbf{LB:} #1}}}
\newcommand{\ym}[1]{\textcolor{yellow}{{\textbf{YM:} #1}}}

\fi

\copyrightyear{2021}
\acmYear{2021}
\setcopyright{acmlicensed}
\acmConference[SIGIR '21] {Proceedings of the 44th International ACM SIGIR Conference on Research and Development in Information Retrieval}{July 11--15, 2021}{Virtual Event, Canada.}
\acmBooktitle{Proceedings of the 44th International ACM SIGIR Conference on Research and Development in Information Retrieval (SIGIR '21), July 11--15, 2021, Virtual Event, Canada}
\acmPrice{}
\acmISBN{978-1-4503-8037-9/21/07}
\acmDOI{10.1145/3404835.3463093}

\begin{document}

\fancyhead{}

\title{A Systematic Evaluation of Transfer Learning and Pseudo-labeling with BERT-based Ranking Models}
\author{Iurii Mokrii}
\authornote{Equal contribution.}
\email{ymokriy@hse.ru}
\affiliation{%
  \institution{HSE University}
  \city{Moscow}
  \country{Russia}
}
\author{Leonid Boytsov}
\authornotemark[1]
\email{leonid.boytsov@us.bosch.com}
\affiliation{
\institution{Bosch Center for Artificial Intelligence}
\city{Pittsburgh}
\country{USA}
}
\author{Pavel Braslavski}
\affiliation{%
  \institution{Ural Federal University}
  \city{Yekaterinburg} 
  \country{Russia} \\
  \institution{HSE University}
  \city{Moscow}
  \country{Russia}
}



\begin{abstract}
Due to high annotation costs making the best use of existing human-created training data is an important research direction. We, therefore, carry out a systematic evaluation of transferability of BERT-based neural ranking models across five English datasets. Previous studies focused primarily on zero-shot and few-shot transfer from a large dataset to a dataset with a small number of queries. In contrast, each of our collections has a substantial number of queries, which enables a full-shot evaluation mode and improves reliability of our results. 
Furthermore, since source datasets licences often prohibit commercial use, we compare transfer learning to training on pseudo-labels generated by a BM25 scorer.
We find that training on pseudo-labels---possibly with subsequent fine-tuning using a modest number of annotated queries---can produce 
a competitive or better model compared to transfer learning.
Yet, it is necessary to improve
the stability and/or effectiveness of the few-shot training,
which, sometimes, can degrade performance of a pretrained model.
\end{abstract}

\ccsdesc[500]{Information systems~Retrieval models and ranking}

\keywords{Neural information retrieval, transfer learning, pseudo-labeling}


\maketitle

\begin{table*}[tb]
  \caption{Effectiveness (MRR) of zero-shot (ZS) and full-shot (FS) transfer}
  \label{tab:zeroandfull}
  \begin{tabular}{lcccccccccc}
    \toprule
    Target $\rightarrow$ & \multicolumn{2}{c}{Yahoo! Answers} & \multicolumn{2}{c}{MS MARCO doc} & \multicolumn{2}{c}{MS MARCO pass} & \multicolumn{2}{c}{DPR NQ} & 
    \multicolumn{2}{c}{DPR SQuAD} \\
    Source $\downarrow$ & ZS & FS & ZS & FS & ZS & FS & ZS & FS & ZS & FS  \\
    \midrule
    Yahoo! Answers & --        & 0.32 & 0.28$^{*}$ & 0.38 & 0.24\ststr          & 0.33 & 0.34\ststr          & 0.49 & 0.47$^{*}$     & 0.65\\
    MS MARCO doc  & 0.15$^{*}$ & 0.33 & --         & 0.38 & \textbf{0.29}$^{*}$ & 0.34 & \textbf{0.44}$^{*}$ & 0.51 & \textbf{0.56}$^{*}$ & 0.65\\
    MS MARCO pass & 0.19$^{*}$ & 0.33 & 0.30\ststr & 0.39 & --                  & 0.34 & 0.43$^{*}$          & 0.49 & 0.52$^{*}$ & 0.65\\
    DPR NQ        & 0.25$^{*}$ & 0.32 & 0.28$^{*}$ & 0.38 & 0.24\ststr          & 0.33 & --                  & 0.49 & 0.53$^{*}$ & 0.64\\
    DPR SQuAD     & 0.24$^{*}$ & 0.33 & 0.24$^{*}$ & 0.38 & 0.22\ststr          & 0.33 & 0.40$^{*}$          & 0.49 & -- & 0.65\\
    \hline
    BM25 & \multicolumn{2}{c}{0.27} & \multicolumn{2}{c}{0.29} & \multicolumn{2}{c}{0.22} & \multicolumn{2}{c}{0.31} & \multicolumn{2}{c}{0.44} \\
    pseudo-labelling & \textbf{0.29} & 0.33 & \textbf{0.31}$^{\#}$ & 0.38 & 0.23 & 0.32 & 0.35$^{\#}$ & 0.48 & 0.50$^{\#}$ & 0.64 \\
  \bottomrule
  \multicolumn{11}{p{0.72\textwidth}}{\textbf{Notes:} Statistically significant differences between pseudo-labeling and BM25 are marked with \#;
  statistically significant differences between transfer learning and pseudo-labeling are marked with $*$.}
\end{tabular}
\end{table*}

\section{Introduction}

A recent adoption of large pretrained Transformer models \cite{vaswani2017attention,devlin2018bert} 
in information retrieval (IR) 
led to a substantially improvement of ranking accuracy compared to traditional, i.e.,
non-neural retrieval models~\cite{craswell2020overview}. 
It also enabled effective zero-shot transfer learning in a monolingual ~\cite{yilmaz2019cross,zhang2020little,althammer2020cross,guo2020multireqa,ruckle2020multicqa} 
and cross-lingual settings \cite{shi2019cross,macavaney2020teaching,althammer2020cross}.

Transfer learning may reduce the need to collect expensive human relevance judgements required for supervised training ~\cite{naturalquestions,han2020crowd}.
However, many source collections such as a popular large-scale MS MARCO~\cite{msmarco}  
have a non-commercial, research-only license, 
which limits practical applicability of transfer learning.
Furthermore, few-shot learning with transferred models may produce results inferior to transfer learning alone \cite{zhang2020little}.
From the methodological point of view, 
prior studies focus primarily on zero-shot and few-shot transfer 
from a dataset with a large number of queries to a dataset with a small number of queries. 
We have also not seen a study that compares transfer learning to 
training of BERT-based models on pseudo-labels  (generated using in-domain data) \cite{dehghani2017neural}.
 
To fill the gap, 
we study transferability of BERT-based ranking models and 
compare transfer learning to training on pseudo-labels generated
using a BM25 scoring function \cite{Robertson2004}. 
We use five diverse English datasets  that differ in terms of document/query  types and/or lengths.
In contrast to previous studies, 
each of our collections has a substantial number of queries,
which enables a full-shot evaluation mode and improves reliability of our results.

Importantly, this short paper focuses on evaluation of existing techniques rather than on improving them.
We ask the following research questions:

\begin{itemize}
    \item\rqone: When training from scratch, how much data does a BERT-based ranker need to outperform BM25?
    \item \rqtwo: Does a model trained on pseudo-labels outperform BM25 (and by how much)?
    \item \rqthree: Is transfer learning always more effective than  BM25?
    \item \rqfour: Is transfer learning more effective than training on pseudo-labels?
    \item \rqfive: Can we improve upon transfer learning and/or pseudo-labeling with a few training examples? 
\end{itemize}

We find that:
 \begin{itemize} 
    \item \rqone: Training a competitive BERT-based models from scratch may require a substantial  number  of annotated queries: From one hundred to several thousands.
    \item \rqtwo: Models trained only on pseudo-labels consistently outperform BM25 by
     5-15\%.
    \item \rqthree: However, transferred models can be worse than BM25.
    \item \rqfour\ and \rqfive: Transferred models are typically better than models trained on pseudo-labels,
    but we can often match or exceed performance of the former by training on 
    a large number of pseudo-labels
    with subsequent fine-tuning using a moderate number of annotated queries.
    \item \rqfive: In that, fine-tuning with a small number of annotated queries can cause a substantial performance
                  degradation, which confirms prior findings \cite{zhang2020little}.
  \end{itemize}
In summary, we find that pseudo-labeling (possibly combined with few-shot training) delivers results competitive with (and sometimes superior to) transfer learning.
However, there is a need to improve the stability and/or effectiveness of the few-shot training.



\section{Related Work}
A detailed discussion of neural ranking models can be found in recent surveys 
\cite{LinDeepIRSurvey2020,mitra2018introduction,guo2020deep}.
The success of early approaches was controversial \cite{lin2019neural},
but the models relying on large pretrained Transformers \cite{vaswani2017attention},
in particular, BERT-based models \cite{devlin2018bert} 
 decidedly outperformed prior neural and traditional models on a variety of \emph{retrieval} tasks
including TREC evaluations \cite{craswell2020overview} and MS MARCO retrieval challenges.\footnote{\url{https://microsoft.github.io/msmarco/}}
Thus, BERT-based rankers are the focus of this study.

A number of studies demonstrated effectiveness of these models in zero-shot transfer learning.
However, unlike our work, they do not explore fine-tuning in a large-data regime and use small test collections,
which may affect reliability of results \cite{UrbanoMM13,BoytsovBW13}.

Specifically, Yilmaz et al.~\cite{yilmaz2019cross} trained a BERT-based model on several collections with short passages 
and tested them on TREC newswire collections. 
By combining BM25 scores with the scores of a BERT-based ranker they outperformed prior approaches.

R{\"u}ckl{\'e} et al.~\cite{ruckle2020multicqa} analyzed transferability 
of BERT-based ranking models trained on questions from 140 StackExchange forums. 
They trained 140 models in a self-supervised fashion 
to retrieve question's detailed description using  a  question title.
They further evaluated 140 models on 9 external collections
and found that BERT-based rankers outperformed traditional IR baselines in most cases.

Several studies employed zero-shot transfer in a cross-lingual setting.
Shi and Lin~\cite{shi2019cross} fine-tuned a multilingual BERT (mBERT)
on the English TREC Microblog collection\footnote{\url{https://trec.nist.gov/data/microblog.html}}
and tested it on Chinese, Arabic, French, Hindi, and Bengali data (each having around 50 annotated topics). 
MacAvaney et al.~\cite{macavaney2020teaching} fine-tuned mBERT on TREC Robust04 and 
tested it on TREC newswire data in Arabic, Chinese, and Spanish (each featuring from 25 to 50 topics).

Although not directly related to our work, zero-shot transfer was evaluated with \emph{traditional}, i.e.,
non-neural, learning-to-rank models.
For example, Macdonald et al. found the transfer to be effective among different variants of ClueWeb collections~\cite{MacdonaldDO15}.

However, not all studies demonstrated the superior transferability of BERT-based models compared to traditional IR baselines. 
Althammer et al.~\cite{althammer2020cross} experimented with BERT-based models trained on legal documents and 
zero-shot transferred them to a patent retrieval task.
Transferred models were at par with BERT-based models trained on in-domain data, however, 
they were outperformed by a BM25 baseline. 
In the study of Thakur et al.~\cite{thakur2021beir} BM25 outperformed transferred BERT-based re-ranking models on six datasets out of 17.
Similarly, in a answer-sentence retrieval task a BM25 scorer combined with BERT subword tokenization
outperformed other methods on five out of eight datasets~\cite{guo2020multireqa}.

Several papers explored the relationship between the amount of training data and the effectiveness of the resulting IR model.
In particular, 
Karpukhin et al.~\cite{karpukhin2020dense} 
showed that increasing the number of training examples gradually improved the quality of a passage retriever.  
Nogueira et al.~\cite{nogueira2020document} observed that T5~\cite{raffel2019exploring} significantly outperformed BERT in a data-poor regime.
In these studies, using more training data always resulted in better performance.
However, Zhang et al.~\cite{zhang2020little} discovered that fine-tuning a BERT-based ranker 
with a few queries on TREC Robust04 collection led to a substantial degradation of performance compared to zero-shot transfer.
This surprising result motivated our \rqfive.
One can train a neural ranker on pseudo-labels
generated by a traditional retrieval model such as BM25 \cite{Robertson2004}.
Although this approach had been shown to be successful in the past \cite{dehghani2017neural},
we are not aware of any recent (and systematic) evaluation of pseudo-labeling with a BERT-based ranker.

\begin{table}[t]
  \caption{Dataset statistics}
  \label{tab:data}
  \begin{tabular}{
  @{\hskip 0.25em}
  l
  @{\hskip 0.6em}p{3em}
  @{\hskip 0.6em}p{3em}
  @{\hskip 0.6em}p{3em}
  @{\hskip 0.6em}p{3em}
  @{\hskip 0.6em}p{3em}
  @{\hskip 0em}
  }
    \toprule
    Dataset & \#queries train &  \#docs & \#relev. /query & \#tok. /query   & \#tok. /doc \\
    \midrule
    Yahoo! Answers & 100K & 819.6K & 5.7 & 11.9 & 63 \\
    MS MARCO doc  & 357K & 3.2M & 1 & 3.2 & 1197 \\
    MS MARCO pass & 788.7K & 8.8M & 0.7 & 3.5 & 75 \\
    DPR NQ &  53.9K & 21M & 7.9 & 4.5 & 141 \\
    DPR SQuAD & 73.7K & 21M & 4.8 & 5 & 141 \\
  \bottomrule
  \multicolumn{6}{p{0.45\textwidth}}{\textbf{Notes:} 
  Development sets have 5K queries, test sets have 1.5K queries. 
  Text length is the \# of \textbf{BERT} word pieces.}
\end{tabular}
\end{table}

\section{Data}

We use five retrieval question-answering (QA) English datasets, whose statistics is summarized in Table~\ref{tab:data}.
Our dataset selection rationale is twofold.
First, we needed a large number of queries for evaluation in different regimes (from zero- to full-shot). 
This was particularly important to answer \rqone.
Second, a representative evaluation requires collections
that differ in terms of document/query  type (e.g., Wikipedia, Web, community QA), 
query types (factoid vs. non-factoid), 
and query/document lengths.

The first dataset---\ttt{Yahoo! Answers}---is the community question answering (CQA) dataset,
which has mostly non-factoid questions.
Users of the service ask questions on virtually any topic while other community members provide answers.
We use a high-quality subset of \ttt{Yahoo! Answers}
created by Surdeanu et al.~\cite{surdeanu2011learning}.\footnote{Collection L6 in Yahoo WebScope: \url{https://webscope.sandbox.yahoo.com}}
We treat all available answers to a question as relevant documents, including answers that are not marked as ``best answers''.
Queries are created by concatenating short questions and their longer descriptions.
We randomly split \ttt{Yahoo! Answers} into training, development, and testing subsets.
We verify that the split has no obvious data leakage,
i.e., that only a small fraction of the questions have duplicates or near-duplicates across splits.

MS MARCO document (\ttt{MS MARCO doc}) and passage (\ttt{MS MARCO pass}) retrieval 
collections are \emph{related} datasets 
created from the MS MARCO reading comprehension dataset \cite{msmarco}
and contain a large number of question-like queries sampled from the Bing search engine log with subsequent filtering.
These queries are not necessarily proper English questions, e.g., ``lyme disease symptoms mood'', 
but they are answerable by a short passage retrieved from a set of about 3.6M Web documents \cite{msmarco}.
Relevance judgements are quite sparse (about one relevant passage/document per query) and 
a positive label indicates that the passage can answer the respective question.
The document retrieval data set (\ttt{MS MARCO doc}) is created by transferring passage-level relevance to original documents
from which passages were extracted \cite{craswell2020overview}.
Thus, a document is considered relevant only if it contains at least one relevant passage.

The DPR data sets were created by Karpukhin et al.~\cite{karpukhin2020dense} by matching Wikipedia passages 
with questions from two reading comprehension data sets: 
Natural Questions \cite{naturalquestions}  and SQuAD v1.1~\cite{RajpurkarZLL16}; we denote respective datasets as \ttt{DPR NQ} and \ttt{DPR SQuAD}. They
processed a Wikipedia dump 
by removing tables, infoboxes, etc., and split pages into 21M passages containing at most 100 words.
We use relevance judgements, questions, and passages provided by the authors.\footnote{\url{https://github.com/facebookresearch/DPR}}

All collections except \ttt{Yahoo! Answers} come with large ``official'' development sets containing at least 5K queries,
a subset of which we used for testing.
Hyper-parameter tuning was carried out on separate sets sampled from the original training data.
For few- and medium-shot training, we randomly sampled training sets of progressively increasing sizes.
Because we had to carry out a large number of experiments, 
we limited the number of samples to three per query set size and used only 1.5K randomly sampled test queries.

\section{Methods}
We use a BM25 scorer \cite{Robertson2004}  tuned on a development set as a main retrieval baseline.
For each query, 100 documents with top BM25 scores are used as an input to a neural re-ranker
as well as to create pseudo-labels \cite{dehghani2017neural}.
Relevant pseudo-labels are created without human supervision by selecting a document  with the highest BM25 score.
We use all available training queries.

We use a 12-layer BERT\textsubscript{BASE} \cite{devlin2018bert,vaswani2017attention} 
with a fully-connected prediction layer as a neural re-ranker \cite{nogueira2019passage}.\footnote{BERT\textsubscript{BASE} performs at par with BERT\textsubscript{LARGE} on MS MARCO  \cite{HofstatterZH20} and thus is a more practical alternative.}
BERT takes a query-document pair as an input.
Long \ttt{MS MARCO doc} documents are truncated to 445 first BERT tokens, but such shortening leads to only small ($\approx1$\%)
loss in accuracy \anonref{boytsov2021exploring}.
Likewise, we keep at most 64 BERT tokens in queries.

The models are trained using a pairwise margin loss (inference is pointwise). 
In a single training epoch,
we select randomly one pair of positive and negative examples per query (negatives
are sampled from 100 documents with highest BM25 scores).
We use an AdamW \cite{loshchilov2017decoupled} optimizer with a small weight decay ($10^{-7})$, a warm-up schedule and a batch size of 16.\footnote{The learning rate grows linearly from zero for 20\% of the steps until it reaches
the base learning rate \cite{Mosbach2020-kn,Smith17} and then goes back to zero (also linearly).}
Note that we use different base rates for the fully-connected prediction head ($2\cdot10^{-4}$) and for the main Transformer layers ($2\cdot10^{-5}$).

We estimated a number of training epochs necessary to achieve good performance when training from scratch. 
To this end, we experimented with a small number of queries on a development set. 
We observe that for all collections, achieving good performance
with only 32 queries required 16-32 epochs. 
We also observe that training for a larger number of epochs may lead to some overfitting, but the effect is quite small (1-3\%).
Thus, we start with 32 epochs for 32 queries and decrease the number of epochs as the training set size increases.
We use this strategy for both training from scratch and fine-tuning a model.

Experiments are carried out using~\flexneuart ~framework. 
Effectiveness is measured using the mean reciprocal rank (MRR),
which is an official metric for MS MARCO data \cite{craswell2020overview}.
For statistical significance testing we use a paired t-test (threshold $=0.01$).

\begin{figure*}[tb]
     \centering
     \begin{tabular}[c]{ccc}
     \begin{subfigure}[c]{0.3\textwidth}
         \centering
         \includegraphics[width=\textwidth]{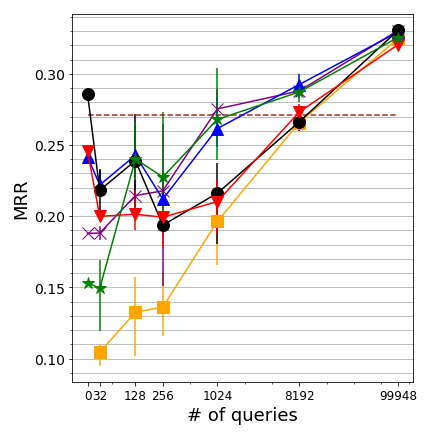}
         \caption{Yahoo! Answers}
         \label{fig:ft_yahoo}
     \end{subfigure}&
     \begin{subfigure}[c]{0.3\textwidth}
         \centering
         \includegraphics[width=\textwidth]{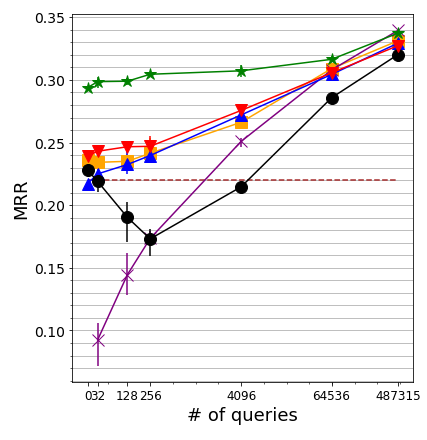}
         \caption{MS MARCO pass}
         \label{fig:ft_msmarco_pars}
     \end{subfigure}&
     \begin{subfigure}[c]{0.3\textwidth}
         \centering
         \includegraphics[width=\textwidth]{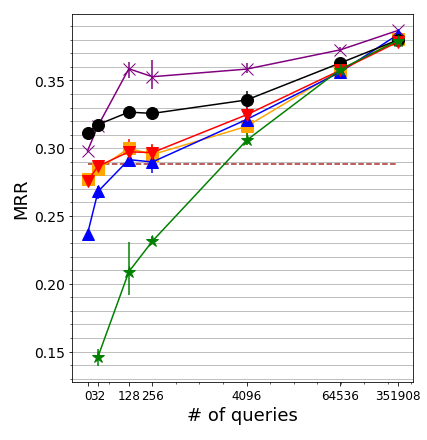}
         \caption{MS MARCO doc}
         \label{fig:ft_msmarco_docs}
     \end{subfigure} \\
     \begin{subfigure}[c]{0.3\textwidth}
         \centering
         \includegraphics[width=\textwidth]{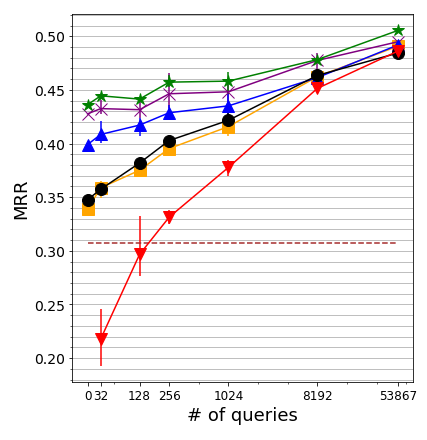}
         \caption{DPR NQ}
         \label{fig:ft_DPR_NQ}
     \end{subfigure}&
     \begin{subfigure}[c]{0.3\textwidth}
         \centering
         \includegraphics[width=\textwidth]{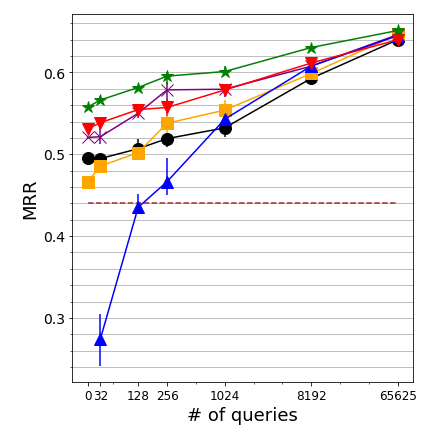}
         \caption{DPR SQuAD}
         \label{fig:ft_DPR_SQuAD}
     \end{subfigure}&
     \begin{subfigure}[c]{0.3\textwidth}
         \centering
         \includegraphics[width=\textwidth]{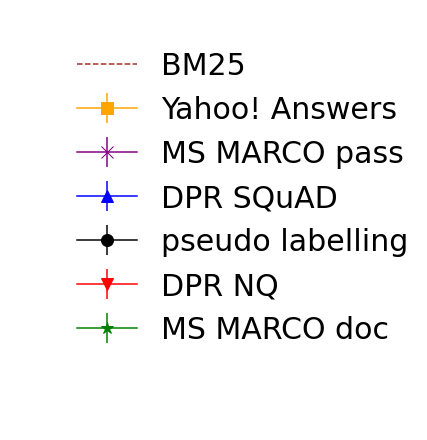}
         \label{fig:ft_legend}
     \end{subfigure}
     \end{tabular}
        \caption{The relationship between MRR and the number of training queries.}
        \label{fig:finetuning}
\end{figure*}

\section{Concluding Discussion of Results}

Table~\ref{tab:zeroandfull} and Figure~\ref{fig:finetuning} contain experimental results.
Figure~\ref{fig:finetuning} shows the relationship between the test accuracy and the training set size
(measured in the number of queries).
Because not all queries have relevant documents (especially in \ttt{MS MARCO pass}),
these sizes are smaller than those in Table~\ref{tab:data}.
Vertical bars indicate the range of test values for training samples of the same size.
Different graphs in a panel correspond to training from a different starting model:
There is graph for training from scratch, from a model trained on pseudo-labels,
as well as one graph per each source collection.

\rqone: we can see that outperforming BM25 requires over 100 annotated queries on DPR data, 
at least 1-2K annotated queries on MS MARCO data and more than 8K annotated queries on \ttt{Yahoo! Answers}.
Judging a single document-pair takes at least one minute on average \cite{han2020crowd,naturalquestions}
and a single query typically needs at least 50 of such judgements \cite{BuckleyDSV07}.
Thus, annotating a single query by one worker can take longer than an hour.
For MS MARCO, this entails several person-months of work just to match the accuracy of BM25.

\rqthree: Transfer learning, however, can be worse than BM25 or outperform it only by a small margin (Table~\ref{tab:zeroandfull}),
which is in line with some prior work \cite{althammer2020cross,guo2020multireqa,thakur2021beir}.
For example, for DPR collections a model transferred from \ttt{Yahoo! Answers} is only $\approx10$\% better than 
BM25. For \ttt{Yahoo! Answers}, \emph{all} transferred models are worse than BM25.
Transfer learning is also mostly \emph{ineffective} on MS MARCO
where only \ttt{MS MARCO doc} model transferred to a related \ttt{MS MARCO pass}
dataset outperformed BM25 by as much as 30\%.

\rqtwo: In contrast, pseudo-labeling \emph{consistently} outperforms BM25
(differences are statistically significant except for \ttt{Yahoo! Answers} and  \ttt{MS MARCO pass}).
Yet, the observed gains (5-15\%) are substantially smaller than those reported by Dehghani et al.~\cite{dehghani2017neural}.

\rqfour\  and \rqfive: 
For almost every source model on DPR and MS MARCO datasets, 
a relatively small  number of annotated queries (100-200)  allow us
to substantially improve upon both the transferred models and models trained on pseudo-labels.
However, we also observe an ``Little Bit Is Worse Than None'' effect \cite{zhang2020little} 
on \ttt{MS MARCO pass} with pseudo-labeling
as well on \ttt{Yahoo! Answers}.

The effect is particularly pronounced on \ttt{Yahoo! Answers}, where few-shot training ruins performance of \emph{every} source model.
We hypothesize that few-shot training can lead to substantial overfitting to a small training set
so that a model ``forgets'' what it learned from source training data.

We believe a similar forgetting happens when the amount of in-domain training data becomes sufficiently large (but it does not have a negative effect).
As the number of training samples increases,
the difference between different pretraining setups decreases:
When we train using all the data, there is virtually no difference between starting
from scratch or from a pretrained model.

To conclude, we note that transferred models are typically better than models trained on pseudo-labels
and these differences are mostly statistically significant (see Table~\ref{tab:zeroandfull}).
However, we can often match or exceed performance of transferred models using a modest number of annotated queries to fine-tune a model trained on \emph{pseudo-labels}.
We thus, hypothesize, that training on pseudo-labels with a subsequent few-shot training on human-annotated data can become
a viable alternative to transfer learning.
Unlike zero-shot models trained on out-of-domain data, this scenario uses only in-domain data.
Thus, it is likely to be less affected by the distribution mismatch between training and testing sets.
However, one needs to improve the stability and effectiveness of the few-shot training, which, nevertheless, is out of the scope of this short paper.



\begin{acks} 
Pavel Braslavski thanks the Ministry of Science and Higher Education of the Russian Federation (``Ural Mathematical Center'' project).
\end{acks}

 \balance

\end{document}